          [1999/12/02 1.01 (PWD)]
\documentclass[a4paper,twocolumn]{esapub} 
\usepackage{natbib}
\usepackage{epsfig}

\title{Gamma-Ray Imaging with the Coded Mask IBIS Telescope}

\author{A. Goldwurm}
\author{P. Goldoni}
\author{A. Gros}
\affil{Service d'Astrophysique /DAPNIA/DSM/CEA - Saclay, 
91191 Gif sur Yvette Cedex, France}

\author{J. Stephen}
\author{L. Foschini}
\author{F. Gianotti}
\affil{ITeSRE/CNR - Via Gobetti 101, 40129 Bologna, Italy}

\author{\\L. Natalucci}
\author{G. De Cesare}
\author{M. Del Santo}
\affil{IAS/CNR , Via del Fosso del Cavaliere, 
00133 Roma, Italy}

\begin{document}
\maketitle
\begin{abstract}
The IBIS telescope onboard INTEGRAL, 
the ESA gamma-ray space mission to be launched in 2002, 
is a soft gamma-ray (20 keV - 10 MeV) device 
based on a coded aperture imaging system. 
We describe here basic concepts of coded masks, 
the imaging system of the IBIS telescope,
and the standard data analysis procedures to reconstruct sky images.
This analysis includes, 
for both the low-energy detector layer (ISGRI) and the high energy layer 
(PICSIT), iterative procedures which decode recorded shadowgrams, 
search for and locate sources, 
clean for secondary lobes, and then rotate and compose sky images.
These procedures will be implemented in the Quick Look and 
Standard Analysis of the INTEGRAL Science Data Center (ISDC) 
as IBIS Instrument Specific Software.
\end{abstract}
\section{Coded Mask Imaging}
Coded aperture systems nowadays find their major application in high
energy astronomy, and in particular in the hard X-ray (3-30 keV) and soft
$\gamma$-ray (30 keV - 20 MeV) 
domains where conventional focusing techniques are
difficult to implement and where the high and
variable background limits the performance of standard on/off monitoring
techniques (see review by Caroli et al. 1987).
\\
In coded aperture telescopes source radiation is spatially modulated 
by a mask of opaque and transparent elements before 
being recorded by a position sensitive detector,
allowing simultaneous measurement of source and background fluxes.
Reconstruction of the sky image
is generally based on a correlation procedure between the recorded image
and a decoding array derived from the mask pattern.
Essential requirements for such systems is that the mask pattern must 
allow each source in the field of view (FOV) to cast a unique shadowgram 
on the detector and that projected shadowgrams must be, as much as possible, 
different to better differentiate the sources.
Assuming a perfect detector (infinite spatial resolution), the 
angular resolution of such a system is then defined by the angle subtended
by one hole at the detector. The sensitive area depends on
the number of all transparent elements of the mask viewed by the detector. 
So, reducing hole size or increasing mask-detector distance 
while increasing accordingly the number of holes 
improves the angular resolution without loss of sensitivity. 
A large field of view can also be obtained since 
the source radiation is modulated for sources which are within an angle 
from the axis given by the arctg of the mask plus detector 
dimensions divided by the mask-detector distance.
To optimize the sensitive area of the detector, masks of dimensions 
greater of or equal to the detector dimensions are employed. 
Two kind of FOV are defined. The fully coded (FC) FOV for which all 
source radiation directed towards the detector plane 
is modulated by the mask and the Partially Coded (PC)
FOV for which only a fraction of this source radiation is modulated, 
while the rest, if detected, cannot be distinguished from the background. 
If holes are uniformly distributed the sensitivity 
is constant in the FCFOV and decreasing in the PCFOV.
\section{Optimum Coded Aperture Systems}
The System Point Spread Function (SPSF) of coded aperture telescopes, 
i.e. the final imaging response to a point source after
reconstruction, depends critically on the mask pattern.
Representing the mask with an array M of 1 (open elements) and 0
(opaque ones), the detector array D will be given by the convolution 
of the sky image S by M plus an unmodulated background array term B,
$$D = S \star M + B$$
Suppose to find a special array M for which exists a
{\it correlation inverse} G such that $M \star G = \delta$-function, 
then we can reconstruct the sky by 
$$S' = D \star G = S \star M \star G + B \star G $$ 
$$~= S \star \delta + B \star G = S + B \star G$$
and $S'$ differs from $S$ only by the $B \star G$ term, 
which for a flat array $B$ is a constant level 
which can be measured and removed.
Such special mask patterns, including those called {\it uniformly redundant
arrays} (URA), were found in the 70s, and they allow the
reconstructed image to be free of secondary lobes (Fenimore \& Cannon 1978).
Most of these patterns are built using binary sets called 
{\it cyclic different sets} which have the remarkable property 
that their cyclic autocorrelation gives a delta function.
The decoding array $G = 2 M - 1$ 
(i.e., G=$+1$ for M=1 and G=$-$1 for M=0) is then a {\it correlation inverse}.
\\
To have a sidelobe-free response a source must be able
to cast on the detector a whole basic pattern (fully coded source).
To make use of all the detector area and to allow more than one source
to be fully coded, the mask basic pattern is normally taken of
the same size and shape of the detector and the total mask made by a cyclic 
repetition ($<$ 2~$\times$~2 for rectangular mask) of the basic pattern.
For such {\it optimum systems} a FCFOV source will always project a 
cyclically shifted version of the basic pattern and correlating 
the detector image with the G decoding array will provide sidelobe-free 
peak with position-invariant shape at the source position.
\\
A source in the PCFOV will instead cast an incomplete pattern 
and its contribution cannot be a-priori subtracted and will produce
secondary lobes (coding noise).
On the other hand the modulated radiation from PC sources can be
reconstructed by extending with a proper normalization the correlation 
procedure to the PCFOV ($\S 4$). 
The complete field of view of the telescope (FOV)
is therefore composed by the central FCFOV of
constant sensitivity and optimum image properties 
(position-invariant and flat sidelobes SPSF) 
surrounded by the PCFOV of decreasing sensitivity and non perfect SPSF.
A source outside the FOV simply contributes to the background level.
\\
These masks also minimize
the statistical errors of the reconstructed peaks.
Since $V = G^2 \star D = \Sigma D $ the variance associated with
each reconstructed sky image pixel is constant in the FCFOV 
and equal to total counts recorded by the detector,
therefore the source signal to noise is simply
$$S/N = { C_S / \sqrt{C_S + C_B}} $$
where $C_S$ and $C_B$ are source and background counts.
These masks also have nearly equal number of transparent and opaque elements 
and therefore offer minimum statistical error in condition of 
high background (typical of the $\gamma$-ray domain).
However the sensitivity also depends on the detector spatial resolution
and an {\it imaging efficiency} factor must be applied to this 
maximum S/N to account for this effect. 
\section {System Point Spread Function}
To perform discrete operations the counts are binned in detector pixels 
and, to avoid loss due to coarse sampling, pixels are of much smaller size 
than the mask elements. 
The correlation can take the form of {\it fine cross-correlation\/},
for which the array
$G$ is itself divided in finer elements with the same sampling
and then correlated \citep{FC81}.
The SPSF is then given by a pyramidal function whose width
(FWHM) is 1 mask element.
However the spatial resolution of real detector is not infinite and
this induces an intrinsic loss in peak reconstruction 
and makes fine sampling (sampling-pixels small) useless.
In this case, and in particular when detector resolution
is not negligible with respect to the mask element size,
it can be shown that the best point-source signal to noise is obtained
by convolving the detector image with the $G$ array convolved by 
the detector point spread function.
For a pixellated detector the blurring function is just a block function 
of width 1 pixel.
Convolving $G$ with this function and performing the correlation 
we obtain a final SPSF which is the convolution of 2 pyramidal functions 
one of width 1 mask element and the other of width 1 pixel. 
\\
Due to finite pixel size, the peak of the SPSF will be 
reconstructed in the best case with an average efficiency given by the 
relation $ (1 - {1 \over 3R})$ \citep{GS95},
where R is the ratio of the mask element size to pixel size.
This gives the average loss in sensitivity when the flux is estimated
by {\it fitting} the image with the SPSF function at fixed source position. 
On the other hands when performing the SPSF deconvolution with given sampling
the reconstructed peak height will be even lower, due to the fact 
that the source will not be at the center of the sampled pixel.
In average this ({\it imaging}) efficiency will be given by 
$(1 - {1 \over 4R})^2$. 
In the case of continuous detectors, like the SIGMA telescope,
the detector blurring can be described by a bi-dimensional Gaussian,
and the SPSF by an analytical function 
(e.g. appendix A in Bouchet et al. 2001), which depends
on the detector spatial resolution (width of the Gaussian).
To precisely evaluate source parameters and their errors 
we compare reconstructed image sectors with the system PSF
by means of the chi-square fitting technique.
\section{Deconvolution and Analysis}
\begin{figure}
\centering
{\epsfig{figure=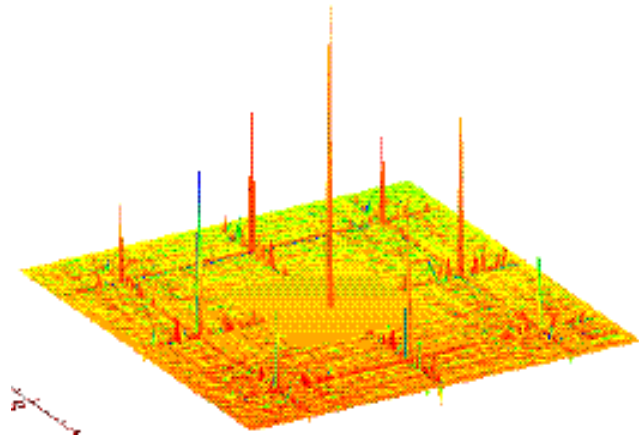 ,width=85mm, height=60mm}}
\caption{SPSF for the IBIS/ISGRI telescope.\label{fig:MuraExfovPsf}}
\end{figure}

Discrete cross-correlation to compute sky and variance images 
can be written
$$ S_{ij}= \sum^{}_{kl} G_{i+k,j+l}D_{kl}  ~~;~~
V_{ij}= \sum^{}_{kl} G^2_{i+k,j+l}D_{kl} $$
where Poisson statistics was assumed.
This standard deconvolution in FCFOV can be extended in the PCFOV
by extending the correlation of the decoding array $G$ with the detector
array D in a non-cyclic form, padding G with 0 elements.
Since only the detector section modulated by the PC source is 
used to reconstruct the signal, the statistical error at the
source position and significance of the ghost peaks are minimized.
However to ensure a flat image in the absence of sources, detector pixels
which for a given sky position correspond to mask opaque elements
must be balanced, before subtraction, with the factor
$b = {n^+ \over n^-}$ where $ n_+ $ is the number of pixels
corresponding
to transparent elements and $ n_- $ to opaque ones for that given sky 
position. This can be written
$$ S_{ij}= \sum^{ }_ kG^+_{i+k,j+l}W_{kl}D_{kl} - B_{ij} 
\sum^{ }_{kl}G^-_{i+k,j+l}W_{kl}D_{kl} $$
where the decoding arrays are obtained from the mask M by $ G^+=M $ and $
G^-= 1-M $,
then padded with 0's outside mask region, and where the sum is performed 
over all detector elements.
In the FCFOV we obtain the same result of the standard cross-correlation.
To consider effects such as satellite drift corrections (see Goldwurm 1995), 
dead areas or other specific conditions, 
a weighting array $ W $ is used to weigh properly
the detector array before correlating it with the $G$ arrays.
The balance array is
$$ B_{ij}=  { \sum^{ }_ kG^+_{i+k,j+l}W_{kl} 
\over \sum^{ }_ kG^-_{i+k,j+l}W_{kl}}  $$
The variance, which is not constant outside the FCFOV, 
is computed accordingly
$$ V_{ij}= \sum D_{kl} \left( G^+_{i+k,j+l}W_{kl}
\right)^2$$
$$+  ~B^2_{ij} \sum D_{kl} \left( G^-_{i+k,j+l}W_{kl} \right)^2 $$
since the cross-terms $G^+ G^-$ vanish.  
Note however that when the weights $ W_{kl} $ refer to the same pixel
in D, the terms $G~W$ must be summed before squaring (see Goldwurm 1995).
The varying effective area can be calculated by similar formula
and used to renormalize, after background subtraction to FCFOV count rates.
All this can be performed for sampling finer than 1 pixel per mask element
and using a G array convolved with detector PSF in order to optimize S/N 
for point sources, with corresponding normalizations.
This procedure can be carried out with a fast algorithm by
reducing previous formulae to a set of correlations
computed by FFT.
\\
The on-axis SPSF on the whole FOV for an optimum system 
(IBIS/ISGRI configuration with MURA mask and R = 2.43, $\S$ 5)
and PSF deconvolution is shown in Fig.~\ref{fig:MuraExfovPsf}. 
Note the peak and flat level in the central FCFOV, 
the secondary lobes (coding noise) in the PCFOV 
and the 8 main ghosts of the source peak in the PCFOV
located at distances, from the source, which are multiple of the 
basic pattern.
\section{IBIS imaging system}
\begin{figure}
\centering
\centerline{\epsfig{figure=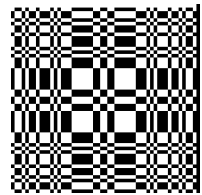 ,width=25mm}}
\caption{The 53 $\times$ 53 MURA basic pattern.\label{fig:Mura}}
\end{figure}

The IBIS coded mask imaging system is composed by a replicated 
Modified URA (MURA) mask of tungsten elements (Fig.~\ref{fig:Mura}) 
and 2 pixellated detector planes of the same size,  
ISGRI for the low energy band (20-1000 keV) and 
PICSIT for the higher band (150 keV - 10 MeV) disposed about 10 cm
below \citep{PU98}.
The MURA \citep{GF89} are nearly-optimum masks and previous discussion
for such system is valid for the IBIS telescope.
The detector planes can be divided in a regular grid of square pixels where
each detector element occupies a pixel of the array. 
Essential imaging characteristics and performances are reported in 
Table~\ref{tab:table1}, including value of efficiency for fitting
procedure and imaging.
\section{IBIS Point Source Location Error}
\begin{figure}
\centering
\centerline{\epsfig{figure=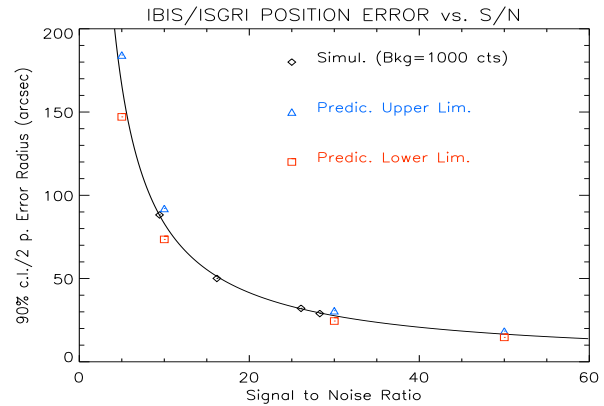 ,width=85mm, height=55mm}}
\caption{
Position errors (at 90$\%$ confidence level in 2 parameters) 
for the IBIS/ISGRI telescope from simulations of a nearly on-axis source at 
different S/N compared to predicted theoretical limits.
Solid line is a function $(S/N)^{-1}$ normalized at S/N=9.5.\label{fig:PSLA}}
\end{figure}

The average Point Source Location Error (PSLE) for an optimum coded aperture
system with a defined System Point Spread Function depends 
on source signal to noise ratio (S/N) as following
$$ PSLE \div {1 \over R~(S/N)} $$
This can be demonstrated by computing, through the covariance matrix, 
the errors on the parameters 
(source position and intensity and background level)
derived by the chi-square fitting of a deconvolved image sector with the SPSF.
We calculated this error, averaged over a large number of uniformly 
distributed source positions, 
for an optimum coded aperture system where mask elements
are an integer number of detector pixels for different values
of mask/pixel sizes and S/Ns.
In Fig.~\ref{fig:PSLA} are reported these theoretical location errors
for mask to pixel ratios R = 2 and 3, 
which are respectively upper and lower limits for the IBIS/ISGRI 
configuration which has an intermediate ratio of $R$ = (11.2/4.6) = 2.43.
\begin{table}
\begin{center}
\caption{IBIS Imaging System and Performances}
\begin{tabular}{ll}
\hline 
MURA basic; total mask     &    53 $\times$~53 ~;~ 95 $\times$~95   \\
Mask element size          &    11.2 $\times$ 11.2 $\times$ 16~$mm^3$\\
ISGRI-Mask distance        &    3200 $mm$ ~$(top-top)$\\
ISGRI active pixels        &    128 $\times$ ~128 \\
ISGRI pix size             &    4. $\times$ ~4. $\times$ ~2.~$mm^3$ \\
ISGRI pixel pitch          &    4.6 $\times$ ~4.6~$mm^2$ \\
ISGRI pixels               &    130 $\times$ ~134 \\
PICSIT-Mask distance       &    3300 $mm$ ~$(top-top)$\\
PICSIT Active Pixels       &    64 $\times$ ~64 \\
PICSIT pix size            &    8.66 $\times$ ~8.66 $\times$ 30 ~$mm^3$ \\
PICSIT pixel pitch         &    9.2 $\times$ ~9.2 ~$mm^2$\\
PICSIT total pixels        &    65 $\times$ ~67  \\
\\
ISGRI  effic. ({\it Fit-Imag}) &    0.86 - 0.81 \\
ISGRI  ang.res.(FWHM)      &    12.0$'$    \\
ISGRI  pixel angle         &    5.0$'$   \\
PICSIT effic. ({\it Fit-Imag}) &    0.73 - 0.63 \\
PICSIT an.res.(FWHM)       &    11.7$'$  \\
PICSIT pixel angle         &    9.6$'$ \\
EXFOV (0 $\%$ sens.)       &    30.6$^\circ$ $\times$ 31.0$^\circ$ \\
FCFOV (100 $\%$ sens.)     &    8.3$^\circ$ $\times$ 8.6$^\circ$ \\
\hline 
\end{tabular}
\label{tab:table1}
\end{center}
\end{table}
In the same plot are reported location errors obtained from analysis of 
simulated observations with IBIS/ISGRI system of an on-axis point source.
200 ISGRI images for a point-like nearly-on-axis source of given S/N
(the S/N is here reduced by {\it imaging efficiency} as defined in $\S 3$) 
are simulated, deconvolved and analyzed to search for the most
significant peak which is then
fitted to the analytical SPSF to derive source position.
Background (uniform distribution) was kept fixed to a value of 1000 cts 
and S/N was made varying 
by reducing input source counts from $\approx$ 2000 to $\approx$ 500.
The standard deviation of the observed offsets between the
best fit positions and the input positions represent then the 1~$\sigma$ 
error in 1 parameter of the estimated position.
These results show that the estimated location error, as expected, 
varies with S/N and R as predicted and can therefore be for bright sources
a small fraction of the angular resolution, 
even for small values of the ratio mask element size to pixel size.
The IBIS/ISGRI telescope, assuming 
no error in pointing axis reconstruction or other systematic effects, can
locate a 30 $\sigma$ point-like source at better than $\pm$ 30$''$.
Absolute error in attitude reconstruction for INTEGRAL 
is expected to be $<$ 20$''$.
\section{Iterative Image Reconstruction}
In a standard analysis, IBIS events or histograms 
are binned in detector images, which are then
corrected for detector and background non-uniformity \citep{AG95} 
and then processed by an iterative algorithm which decodes,
cleans and composes sky images.
For each detector image a sky image and its variance are obtained
using the deconvolution procedure of $\S 4$,
and then iteratively searched for sources and cleaned of the source side
lobes.
In this iterative process the source peaks 
are fitted with the proper system PSF
and finely located. Then the source contribution
to the image is modeled in detail and subtracted.
The images are rotated, projected and summed after beeing
weighted with the variance, and then searched for further contributions.
Note that the search for significant excesses must be performed taking 
into account that these are {\it correlation images} and
the critical level at which an unknown excess is significant 
increases typically from 3~$\sigma$ to 5~-~6~$\sigma$ \citep{C87}.
This procedure was used in simulations to assess capability of IBIS/ISGRI
telescope for a number of specific scientific cases 
(see e.g. 
Goldwurm et al. 1999, 2000, and Goldoni et al. 2000 these proceedings),
and it is now being implemented in software modules to be integrated
into the ISDC scientific analysis pipelines.
%

%
%
\end{document}